\documentclass[twocolumn,superscriptaddress,floatfix,preprintnumbers]{revtex4}
\usepackage{graphics,amssymb,amsmath,epsfig,color}

\usepackage{graphicx}
\usepackage{dcolumn}
\usepackage{bm}

\usepackage[utf8]{inputenc}
\usepackage[T1]{fontenc}
\usepackage{mathptmx}

\usepackage{multirow} 
\usepackage{lineno} 

\newcommand{\tc}{T$_{\mathrm{C}}$}

\begin{document}

\title{Ferromagnetic order of ultra-thin La$_{0.7}$Ba$_{0.3}$MnO$_3$ sandwiched between SrRuO$_3$ layers}

\author{Cinthia Piamonteze}
\email{cinthia.piamonteze@psi.ch}

\affiliation{Swiss Light Source, Paul Scherrer Institut, CH-5232 Villigen PSI, Switzerland}

\author{Francis Bern}
 \altaffiliation[Also at ]{Laboratoire de Physique de la Mati\`ere Condens\'ee, \'Ecole Polytechnique, Palaiseau F-91128, Palaiseau cedex, France} 
\affiliation{Felix-Bloch-Institut f\"ur Festk\"orperphysik Universit\"at Leipzig D-04103 Leipzig, Germany}

\author{Sridhar Reddy  Venkata Avula}
\affiliation{Swiss Light Source, Paul Scherrer Institut, CH-5232 Villigen PSI, Switzerland}

\author{Michal Studniarek}
\affiliation{Swiss Light Source, Paul Scherrer Institut, CH-5232 Villigen PSI, Switzerland}

\author{Carmine Autieri}
 \altaffiliation[Also at ]{Consiglio Nazionale delle Ricerche CNR-SPIN, UOS Salerno, 84084 Fisciano (Salerno), Italy}
\affiliation{International Research Centre MagTop, Institute of Physics, Polish Academy of Sciences, Aleja Lotnik\'ow 32/46, PL-02668 Warsaw, Poland}

\author{Michael Ziese}
\affiliation{Felix-Bloch-Institut f\"ur Festk\"orperphysik Universit\"at Leipzig D-04103 Leipzig, Germany} 

\author{Ionela Lindfords-Vrejoiu}
\affiliation{II. Physikalisches Institut Universit\"at zu K\"oln D-50937 K\"oln, Germany}

\begin{abstract}
We demonstrate the stability of ferromagnetic order of one unit cell thick optimally doped manganite (La$_{0.7}$Ba$_{0.3}$MnO$_3$, LBMO) epitaxially grown between two layers of SrRuO$_3$ (SRO) by using x-ray magnetic circular dichroism. At low temperature LBMO shows an inverted hysteresis loop due to the strong antiferromagnetic coupling to SRO. Moreover, above SRO \tc\ the manganite still exhibits magnetic remanence. Density Functional Theory calculations show that coherent interfaces of LBMO with SRO hinder electronic confinement and the strong magnetic coupling enables the increase of the LBMO \tc. From the structural point of view, interfacing with SRO enables LBMO to have octahedral rotations similar to bulk. All these factors jointly contribute for stable ferromagnetism up  to 130~K for a one unit cell LBMO film.
\end{abstract}

\maketitle


Optimally doped manganite (La$_{0.7}$Sr$_{0.3}$MnO$_3$ - LSMO) has attracted interest for use in magnetic tunnel junctions due to its high values of spin polarization and  Curie temperature (T$_C$) \cite{Park:1998vu}. However,  such applications have been partially hindered, due  to findings that \tc\ strongly decreases for ultra-thin layers, with a non-ferromagnetic insulator layer of about 5 unit cells \cite{Bibes:2001gk,Huijben:2008go,Chen:2019fg}. Several reasons have been attributed for the origin of the magnetic dead layer in manganites. Among them, charge transfer\cite{Yamada:2004wv}, octahedral rotation\cite{EJMoon:2014fo} and symmetry breaking\cite{SValencia:2014} are likely to play a role.

 On the other hand superlattices of LSMO with SrRuO$_3$ (SRO) or La$_{0.7}$Sr$_{0.3}$CrO$_3$ exhibit ferromagnetism for single LSMO layers down to 2 unit cells (u.c.), corresponding to around 0.8\,nm \cite{Koohfar:2020fl,Ziese:2012gb}. In superlattices composed of antiferromagnetic layers of manganite (La$_{2/3}$Ca$_{1/3}$MnO$_3$) and ruthenate (CaRuO$_3$) a ferromagnetic metallic ground state was observed, and attributed to charge transfer at the interface \cite{Chen:2013bt}.  LSMO and SRO couple antiferromagnetically via the interfacial oxygen 2p states\cite{Lee:2008et} and heterostructures of manganites and ruthenates exhibit a complicated antiferromagnetic structure as a function of field and temperature.\cite{Huang:2018jz}. In fact, superlattices combining manganites and ruthenates  were  proposed candidates for synthetic antiferromagnets\cite{Chen:2017wm}.

Here we investigate the stability of ferromagnetic order in one \,u.c.-thick (u.c.~=~unit cell) La$_{0.7}$Ba$_{0.3}$MnO$_3$ (LBMO) interfaced epitaxially with two layers of 3\,u.c.-thick SrRuO$_3$ (SRO), grown on SrTiO$_3$ (001) (STO), which will be called 3|1|3 from now on. Using x-ray magnetic circular dichroism (XMCD) we measure element specific magnetization curves, thus being able to investigate the magnetism of LBMO and SRO separately. We show that at low temperature LBMO shows an antiferromagnetic coupling to SRO, as also observed in superlattices through total magnetometry\cite{Ziese:2010ke}. Interestingly, our data shows that LBMO still exhibits magnetic remanence, even above SRO T$_C$. To get more insight in the magnetic properties of the 3|1|3 heterostructure, we perform Density Functional Theory (DFT) calculations. Our calculations show that a combination of electronic and atomic structure together with the strong magnetic coupling between SRO and LBMO  help stabilizing ferromagnetism in ultra-thin LBMO.

\begin{figure}[tb]
\includegraphics[width=0.45\textwidth]{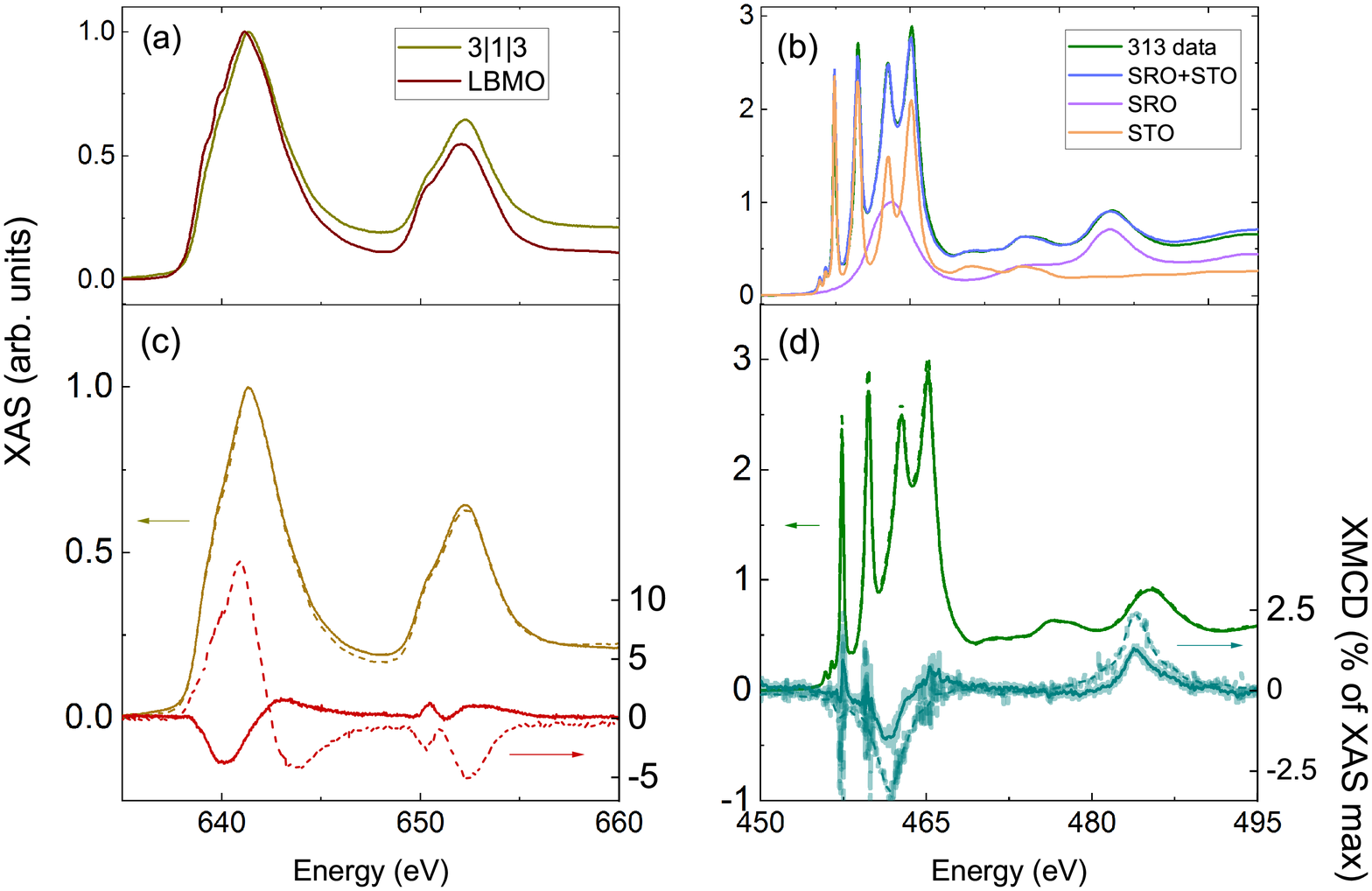}
  \caption{(a) Mn XAS for 3|1|3 compared to LBMO single layer. (b)Ru XAS measured for the 3|1|3 trilayer (blue) compared to a simulation (blue) of the spectra for 3|1|3 using a combination of the measured data for SRO and STO. The SRO (violet), STO (orange) contributions to the simulated XAS are also shown. The data for 3|1|3 are normalized such that the maximum of SRO contribution is at 1.  (c) Mn and (d) Ru XAS (left scale in arbitrary units) and XMCD (right scale in \%\ of the XAS maximum) spectra measured at 10~K and 6.8~T. The continuous lines correspond to measurements in GI and the dashed lines in NI.}
  \label{fig:xmcd}
\end{figure}

High angle annular dark field imaging and electron energy loss spectroscopy performed with a Cs-corrected scanning transmission electron microscope showed the sharp interface of the systems studied here\cite{Bern:2016ev}. The 3|1|3 heterostructure was previously investigated by anomalous Hall effect and SQUID magnetization\cite{Bern:2016ev}. The XAS and XMCD spectra for Ru and Mn are shown in figure \ref{fig:xmcd}.  The measured XAS for Mn in 3|1|3 is in agreement with other published spectra from optimally doped manganites and very similar to the one we have measured for a 30\,nm of LBMO (fig. \ref{fig:xmcd}(a)). No  contribution from Mn$^{2+}$ is seen, which often is visible in ultra thin  layers directly grown on STO\cite{Lee:2010go}. The Ru M$_3$ edge overlaps with Ti L$_{3,2}$ edges. Figure \ref{fig:xmcd}(b) shows the comparison of the Ru XAS in 3|1|3 with the one measured for a 30\,nm thick SRO film and the Ti L$_{3,2}$ measured in a STO crystal. The measured 3|1|3 Ru XAS can be very well reproduced by a combination of the measured SRO and STO spectra on the same energy range, as shown in figure \ref{fig:xmcd}(b). 

Figures \ref{fig:xmcd}(c) and (d) show the XAS and XMCD measured for Ru and Mn, respectively, measured at 10\,K and 6.8\,T for 3|1|3 heterostructure. The XMCD signal is proportional to the net magnetic moment projected along the x-ray beam direction. Therefore, two measurement geometries are used for probing different magnetization directions. Normal incidence (NI) measurement probes the out-of-plane axis while grazing incidence measurement (GI) probes predominantly the component of in-plane magnetization.  In NI (dashed lines in figure \ref{fig:xmcd}(c),(d)) Mn and Ru have opposite sign for the XMCD signal evidencing the antiferromagnetic coupling between these two layers. Since SRO has a larger contribution to the total magnetization, the Ru moment is parallel to the field while Mn is opposite to the applied field. The Ru XMCD is about two times larger in NI than in GI, which agrees with the expected out-of-plane easy axis measured in SRO films deposited on STO(001)\cite{Ziese:2010ep}. On the other hand, the Mn XMCD in GI has the same sign as Ru showing that at high magnetic field both have a component parallel to the applied field. The overall smaller XMCD signals for GI indicates a canted magnetization state.

\begin{figure}
\includegraphics[width=0.45\textwidth]{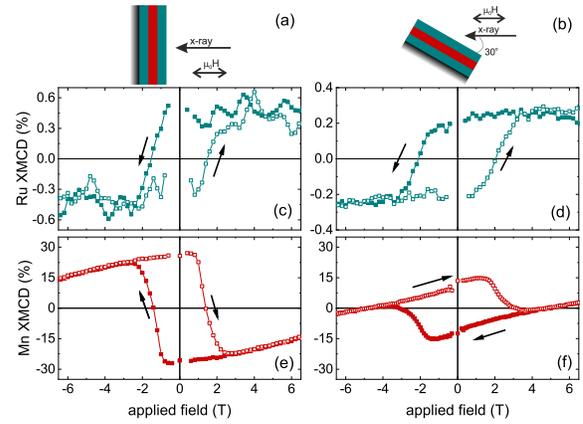}
  \caption{(a) Measurement geometry sketch for normal and (b) grazing incidence, probing out-of-plane and in-plane magnetization, respectively. (c) SRO hysteresis measured in normal and (d) grazing incidence.   (e) LBMO hysteresis measured in normal and (f) grazing incidence. All measurements were performed with the sample at 10\,K.}
  \label{fig:hyst}
\end{figure}

In order to understand further the field dependence of the individual layers in both geometries, we have measured the XMCD signal as a function of applied field in order to obtain an element specific hysteresis curve. Figure \ref{fig:hyst} shows the magnetization curves measured at Ru  (figure \ref{fig:hyst} (c) and (d))  and Mn  (figure \ref{fig:hyst} (e) and (f)) resonances as a function of applied magnetic field. Figures \ref{fig:hyst}(c) and (d) show further confirmation for the out-of-plane easy axis in these trilayers: the Ru XMCD signal is larger and the coercive field smaller for the NI measurement compared to GI. The coercive field measured for out-of-plane (figure~\ref{fig:hyst}(c)) is $\approx$~2.0\,T. This is about twice the value measured for a bare 5\,nm-thick ($\approx$~12 u.c.) SRO film\cite{Ziese:2010ep}. This difference likely comes from the larger contribution of the surface anisotropy in the much thinner SRO layer investigated here as well as a reduced demagnetization field due to the AF configuration. As mentioned before, at GI the largest contribution is from the in-plane magnetization, but an out-of-plane component also contributes to the signal.

The Mn magnetization for the 1 u.c.-thick LBMO for NI (fig.~\ref{fig:hyst}(e)) shows a clear inverted hysteresis, evidencing again the antiferromagnetic coupling between optimally doped manganite and SRO \cite{Lee:2008et,Ziese:2012gb}. Similar inverted hysteresis were measured for thicker LSMO/SRO bilayers using XMCD \cite{Das:2019ks}. Figure ~\ref{fig:hyst}(e) shows that a single LBMO layer still exhibits ferromagnetic behavior at 10~K.  In GI (figure \ref{fig:hyst}(f)), the LBMO layer does not rigidly oppose the SRO magnetization, as in NI. Instead, around 5\,T  the LBMO film XMCD is close to zero. Above this applied field the Mn magnetization changes sign having a component in the direction of the applied magnetic field.   This shows that the Mn-Mn double exchange coupling and the in-plane magnetic  anisotropy for  LBMO together with the Zeeman energy overcome the antiferromagnetic coupling between Mn and Ru. This is particularly easier at grazing incidence since the Ru magnetic moment component along the field direction is smaller. A quantitative estimation of the magnetic moment is obtained by applying sum rules\cite{Thole:1992tm,Carra:1993uz} to the XMCD spectra at applied field and in remanence (see SI). For Mn the moments found for 3|1|3 are $1.8(2)\mu_B$ (NI) and $-0.33(5)\mu_B$ (GI) at 6.8~T;  3.1(5)$\mu_B$ (NI) and 1.9(3)$\mu_B$ (GI) at remanence. For LBMO single film the moment found was 3.4 $\mu_B$. For the SRO single film the moment found was 1.37(7)$\mu_B$, in agreement with neutron studies\cite{Lee:2013iw}. The Ru moment in SRO was probed by XMCD with a certain disparity in results \cite{Agrestini:2015km,Okamoto:2007dd}. The sum rules on the 3|1|3 Ru data turned out to have very large error bars due to the uncertainty of the XMCD baselines in comparison to the magnitude of the signal. For this reason we scaled the Ru XMCD for the 3|1|3 to the one for the SRO single film for an estimate of the moment size. The Ru XMCD in 3|1|3   is about 75\%\ of the one in SRO at NI and 40\%\ in GI (see SI). 

\begin{figure}[tb]
\includegraphics[width=0.45\textwidth,angle=0]{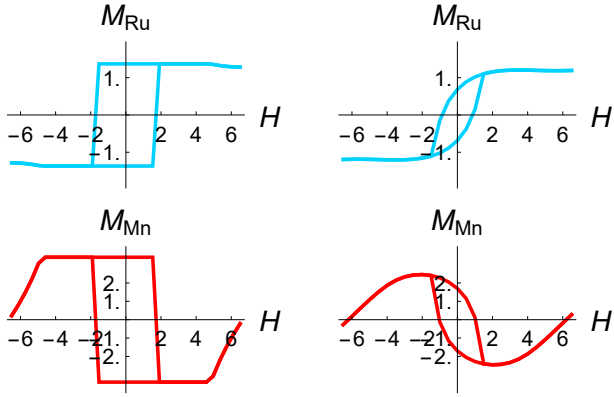}
  \caption{Same quantities as Fig. \ref{fig:hyst} using the theoretical results of the semiclassical model. The unit of the magnetization is in $\mu_B$, the unit of the magnetic field is in Tesla. Left  hand side graphs are for NI and right hand side graphs are for GI.}
  \label{fig:semiclassical}
\end{figure}

We model the hysteresis using the magnetic total energy in the semiclassical form of the 6 SRO layers coupled with 1 LBMO layer. We consider the magnetic exchange between the Ru and Mn atoms, the magnetocrystalline anisotropy and the interaction between the spin and the magnetic field.
\begin{multline}
E=2J^{001}_{Ru-Mn}\cos(\theta_{Ru}-\theta_{Mn})+\\
+6K_{Ru}\cos^2(\theta_{Ru})+K_{Mn}\cos^2(\theta_{Mn})+\\
-6M_{Ru}H\cos({\theta_{Ru}-\theta_H})-M_{Mn}H\cos({\theta_{Mn}-\theta_H})
\end{multline}
where the magnetization of the Ru and Mn atoms are fixed to the experimental values M$_{Ru}$=1.37 $\mu_B$ and M$_{Mn}$=3.4 $\mu_B$, while K$_{Ru}$ and K$_{Mn}$ are the magnetocrystalline anisotropy for the Ru and Mn spins, respectively.
The angles $\theta_{Ru}$ and $\theta_{Mn}$ are the angles of the spins with respect to the reference system (the film surface in our case). Because of the AFM coupling between Ru and Mn, the $\theta_{Ru}$ and $\theta_{Mn}$ angles differ by 180 degrees at zero magnetic field and they change with the magnetic field.   
H and $\theta_H$ are the intensity and the angle of the magnetic field, in the experimental setup $\theta_H$=$\frac{\pi}{2}$ and $\frac{\pi}{6}$. We tune the field H, and we calculate $\theta_{Ru}$ and $\theta_{Mn}$ for the given magnetic field from the minimum of the total energy.
For H larger than the coercive field, the Ru moment aligns to the magnetic field and $\theta_{Ru}$ becomes equal to $\theta_H$.
There is a competition between the 6 layers of SRO and the single layer of LBMO. Since $6m_{Ru}>m_{Mn}$, the dominant behaviour is given by the magnetization of the 6 layers of SRO that follow the magnetic field, as a consequence the LBMO aligns antiparallel to the magnetic field.
The competition between $4J^{001}_{Ru-Mn}$ and $M_{Mn}H$ decides the rotation of the Mn layer.
The result is displayed in  Fig. \ref{fig:semiclassical}, showing a good agreement with the experiment. The calculations consider a single domain and that could explain the discrepancy of the magnetization for Ru measured in GI. In the experiment clearly not all domains align with the applied field, while in the calculations the single domain does.

\begin{figure}
\includegraphics[width=0.45\textwidth]{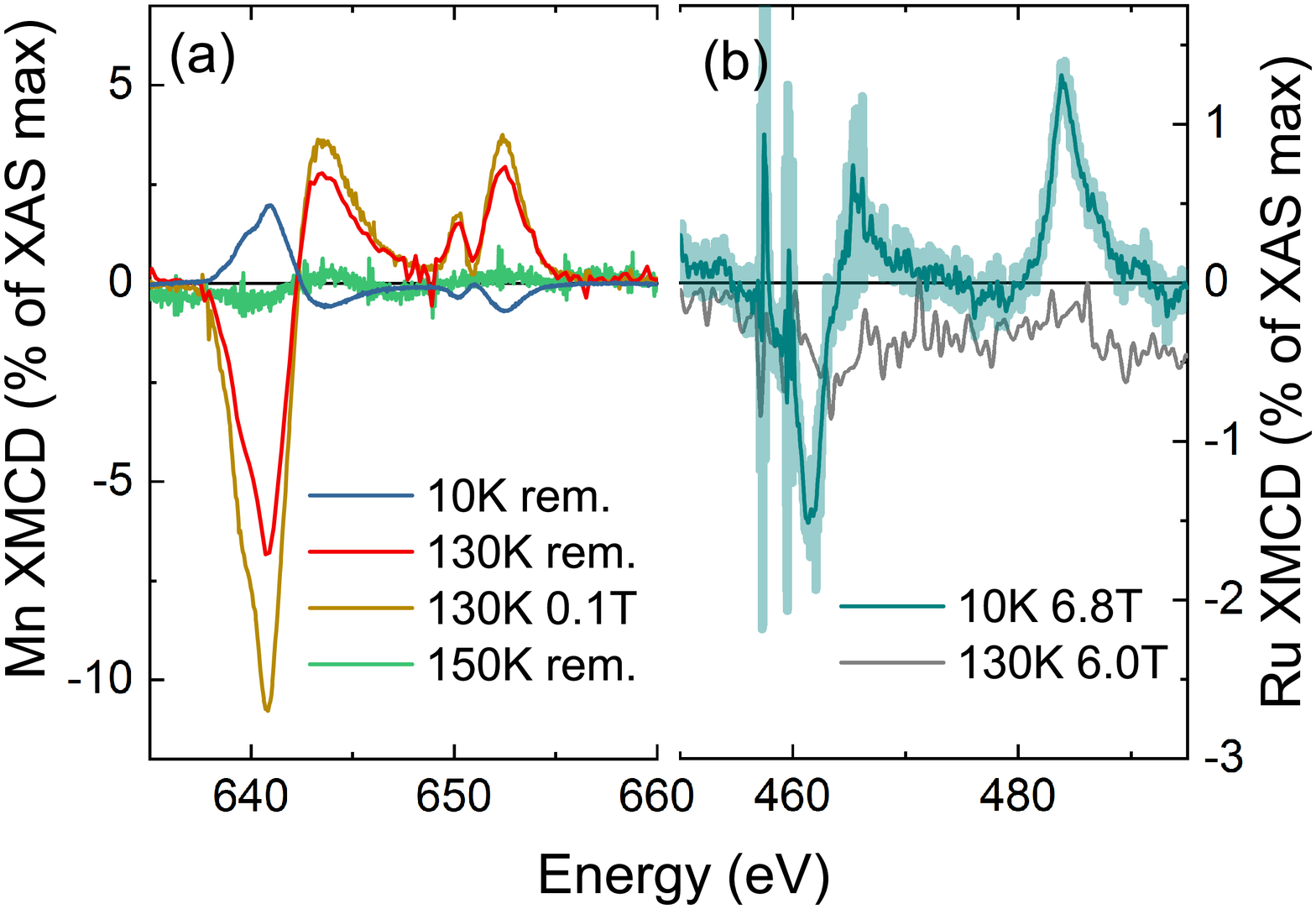}
  \caption{XMCD measured in grazing incidence at (a) Mn L$_{2,3}$ edges and (b) Ru M$_{2,3}$ edges. Temperature and applied field are indicated in the legend. Remanence (rem.) measurements were performed at no applied field after saturation at 6.0\,T.}
  \label{fig:xmcdtemp}
\end{figure}

Next we look at the temperature dependence of the magnetic behaviour.
Figure \ref{fig:xmcdtemp} shows the XMCD data measured in GI at remanence for LBMO at 130\,K and 150\,K compared to 10\,K. The data for SRO at 130\,K and applied field is  plotted for reference. For technical reasons,  it is very difficult to have good signal/noise in the magnetization curve for fields close to zero, making it a  challenge to detect hysteresis opening below $\approx$~50\,mT. For this reason, we choose to measure the XMCD in remanence as an evidence for the presence or not of ferromagnetic order. The remanence data are measured at no applied field after saturating the moments at 6\,T. The superconducting magnet coil used has remanent field of approximately 10\,mT.  At 10\,K (blue curve in figure \ref{fig:xmcdtemp}) the remanence signal for LBMO is opposite to  Ru, as expected from the XMCD $vs.$ field data shown in figure \ref{fig:hyst}. At 130\,K,  XMCD signal for Ru is below the detection level, even at 6\,T. This is not so surprising since the T$_C$ for SRO in these trilayers is around 100\,K, as shown by Bern \textit{et~al} \cite{Bern:2016ev}. The  XMCD signal for LBMO at 0.1\,T and 130\,K has the opposite sign as for LBMO at 10\,K, showing that at this temperature the XMCD for LBMO is parallel to the applied magnetic field. This is an additional evidence that, indeed, SRO is not anymore ferromagnetic and LBMO acts as an independent magnetic layer. When removing the applied field, the Mn XMCD keeps the same sign and is reduced to 67\,\%\ of the value at 0.1\,T showing a clear magnetic remanence. Therefore, the XMCD data unequivocally shows that even above the T$_C$ for SRO, the single LBMO layer still retains its ferromagnetic ordering. When increasing the temperature to 150\,K, the XMCD  signal for LBMO is not anymore detectable as shown by the green curve in figure \ref{fig:xmcdtemp}(a).

\begin{figure}[tb!]
\includegraphics[width=0.45\textwidth,angle=0]{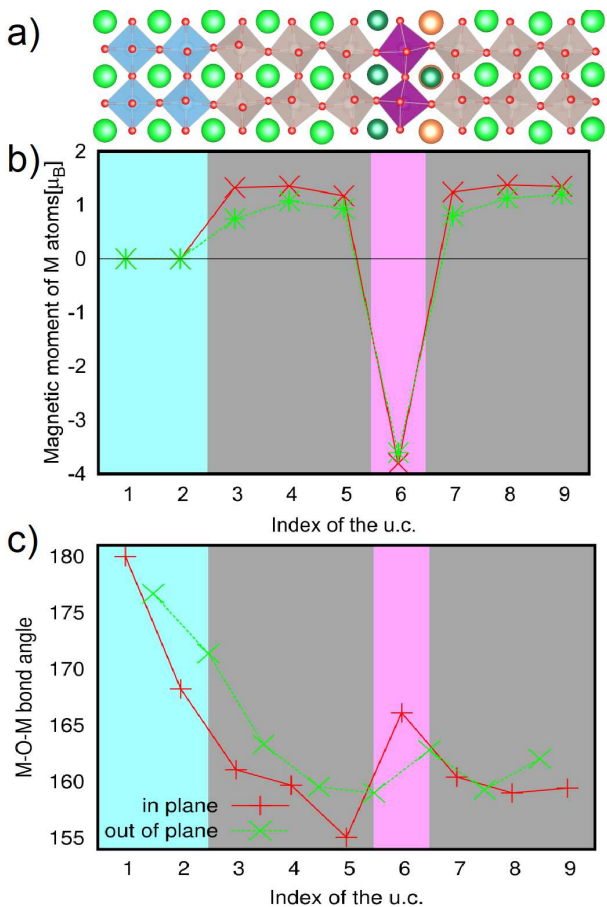}
\caption{a) Crystal structure of the 3|1|3 heterostructures obtained after structural relaxation in DFT. We use the same color of the atoms to define the corresponding regions in the bottom figures.
b) Magnetic moments of the metal atoms in DFT for U$_{Ru}$=0.2 eV and U$_{Mn}$=3 eV (green dashed line) and U$_{Ru}$=1 eV and U$_{Mn}$=6 eV (red solid line). The connecting lines are a guide for the eyes.
c) In plane and out of plane M-O-M bond angles of the 3|1|3 hetetrostructure in DFT for U$_{Ru}$=1 eV and U$_{Mn}$=6 eV. The lines are a guide for the eyes.}\label{fig:DFT_MOM}
\end{figure}

From the DFT calculations, the magnetic configuration of the ground state is represented by the Mn-spins antiparallel to the Ru-spins. 
The magnetic profile is reported in Fig. \ref{fig:DFT_MOM}b) for two sets of Coulomb repulsion. In the first set, we have used the values in the bottom of the typical range (U$_{Ru}$=0.2 eV, U$_{Mn}$=3 eV) while in the second set we have used values in top of the typical ranges (U$_{Ru}$=1 eV, U$_{Mn}$=6 eV).\cite{Roy15,Autieri:2014_NJP,Keshavarz:2017_PRB}
In both cases, we find the largest magnetic moment for the SRO in the inner layers. The average magnetic moment is in the range 0.9-1.3 $\mu_B$ for the Ru and 3.6-3.8 $\mu_B$ for the Mn; these quantities are strongly dependent on the Coulomb repulsion. Lower values of U$_{Mn}$ will make the theoretical value closer to the 3.4 $\mu_B$ experimentally found for the LBMO. 
The increase in the T$_{C,LBMO}$ due to the presence of the SRO is estimated in mean field approximation as $\frac{J_{Mn,Ru}^{001}}{2J_{Mn,Mn}^{100}}$, which is of the order of 0.08-0.09.
This Ru-Mn magnetic coupling produces an increase of 8-9\% of the T$_{C,LBMO}$ with respect to an isolated 1 u.c. of LBMO, in line with the experimental results. 
Additional contribution to the T$_{C,LBMO}$ could come from the increase in dimensionality. This is indicated by the density of states (see SI), which show that the Ru and Mn bandwidths lie in the same energy range, avoiding the quantum confinement.

We have also looked at octahedral rotations of the 3|1|3 heterostructure. STO with its cubic structure has no octahedral rotation and will likely inhibit the corner sharing octahedral rotation in LBMO.  DFT results in Fig. \ref{fig:DFT_MOM}c)  show how the octahedral distortions behave  for the 3|1|3 heterostructure. The in plane M-O-M bond angle in the first layer of STO is theoretically constrained to be 180 degrees. The STO suppresses the octahedral rotations of the layers interfaced with it, but going away from  STO the octahedral rotations increase. Despite the large octahedral rotations of the SRO, we can observe that in the LBMO region the octahedral rotations are  comparable with bulk values of  LBMO. Therefore, the SRO prevents the reduction in critical temperature via structural effects.

In summary our results show that one u.c. thick LBMO has a \tc between 130\,K and 150\,K when epitaxially interfaced with two adjacent 3 u.c. thick SRO layers. This shows  greatly improved ferromagnetic properties compared to a bare ultra-thin film of optimally doped manganite.  DFT calculations show that  interfacing with SRO adjacent layers provides a 3D electronic structure to the LBMO, hindering quantum confinement effects. The strong Ru-Mn magnetic coupling also enhances LBMO \tc\ even when SRO is already in the paramagnetic phase. In addition SRO favorably acts like a buffer that enables LBMO octahedral rotation close to  bulk values. All these effects combined contribute to the stable ferromagnetic state for LBMO. The results reported here demonstrate how impactful epitaxial growth is for the physical properties of perovskite oxides and that effective engineering of the properties can be obtained by suitable choice of the substrate and buffer layers. We  found a particular solution for the design of  ferromagnetically stable ultra-thin  epitaxial films, showing that there exist possibilities to circumvent the notorious dead layer effect that has been thought to annihilate the ferromagnetic order in ultra-thin manganite layers.

\begin{acknowledgements}
C. A. is supported by the Foundation for Polish Science through the International Research Agendas program co-financed by the European Union within the Smart Growth Operational Programme.
C. A. acknowledges the access to the computing facilities of the Interdisciplinary
Center of Modeling at the University of Warsaw, Grant No. G73-23 and G75-10. C. A. acknowledges the CINECA award under the IsC81 "DISTANCE" Grant for the availability of
high-performance computing resources and support. I.L.-V. thanks Gennady Logvenov and Georg Cristiani for the use of the PLD system for the sample fabrication. S. R. V. A. thanks funding from the Swiss National Science Foundation, grants  2000-0\_192393 and 200021\_169467.
\end{acknowledgements}


\pagebreak
\widetext
\begin{center}
\textbf{\large Supplementary Material- \\ Ferromagnetic order of ultra-thin La$_{0.7}$Ba$_{0.3}$MnO$_3$ sandwiched between SrRuO$_3$ layers}
\end{center}
\setcounter{equation}{0}
\setcounter{figure}{0}
\setcounter{table}{0}
\setcounter{page}{1}
\makeatletter
\renewcommand{\theequation}{S\arabic{equation}}
\renewcommand{\thefigure}{S\arabic{figure}}

\section{Trilayer growth}

\begin{figure}[hb]
\includegraphics[width=0.45\textwidth]{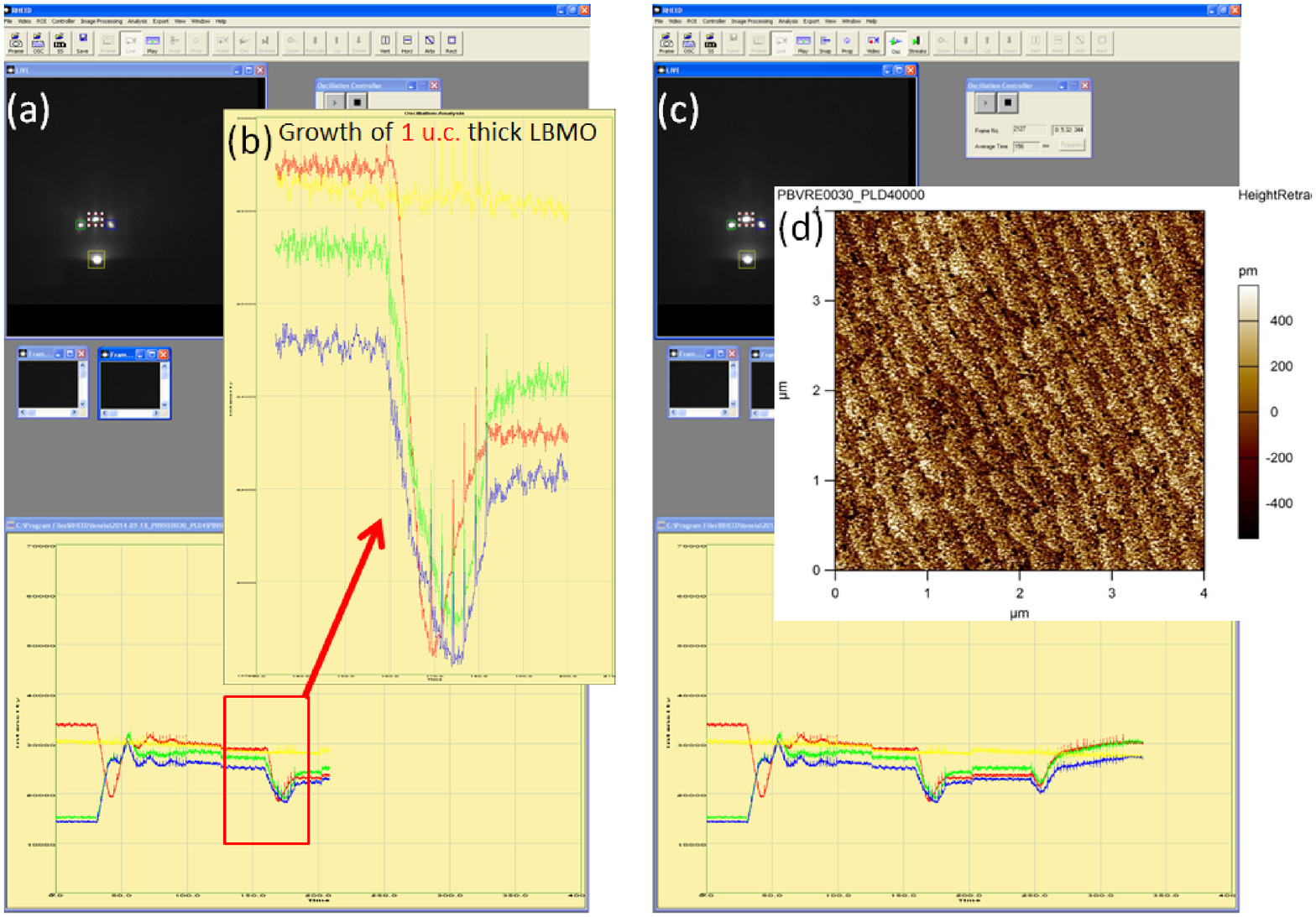}
  \caption{(a) The RHEED pattern and the RHEED signal as a function of time during the deposition of the bottom SRO and middle LBMO layer. (b) The zoom out  of the RHEED intensity as a function of time for the growth of the LBMO layer that is one unit cell thick. (c) The RHEED pattern and the RHEED signal as a function of time during the deposition of the bottom SRO and middle LBMO layer and the topmost SRO layer. (d) An AFM micrograph (4 µm x 4 µm area) of the as-grown trilayer immediately after fabrication. }
  \label{fig:RHEEDtrilayer}
\end{figure}

We monitored the growth of the the 3|1|3 trilayer by in situ reflective high-energy electron diffraction (RHEED). The layers were grown by pulsed-laser deposition on the SrTiO$_3$(100) substrate heated at 973 K in an atmosphere of 0.113 Torr O$_2$ pressure.  Figure \ref{fig:RHEEDtrilayer} shows a summary of the RHEED data and an atomic force micrograph that was taken on the surface of the as-grown sample. The RHEED intensity of the specular spot showed one complete oscillation after 45 laser pulses shot at a repetition rate of 2 Hz, while depositing the LBMO layer (see details in Fig. \ref{fig:RHEEDtrilayer}(b)) . This is a strong indication that one unit cell thick LBMO was grown. The AFM topography image (Fig. \ref{fig:RHEEDtrilayer}(d)) shows a smooth surface with uniform terraces that come from the vicinal substrate.

\section{Computational details}
We perform spin polarized first-principles density functional calculations within the Local Spin Density Approximation (LSDA) by using the plane wave VASP DFT package\cite{VASP,VASP2} and the Perdew-Zunger parametrization\cite{Perdew81} of the Ceperley-Alder data\cite{Ceperley80} for the exchange-correlation functional. 
The choice of LSDA exchange functional is suggested by the literature where the Generalized Gradient Approximation was shown to perform worse than LSDA for SrRuO$_3$ overestimating the magnetization.\cite{Autieri:2016gk} The interaction between the core and the valence electrons was treated with the projector augmented wave (PAW) method\cite{Blochl94} and a cutoff of 450 eV was used for the plane wave basis. The computational unit cells are constructed as supercells with two SrTiO$_3$ layers having 20 {\AA} of vacuum. 
The manganite/ruthenate interface was reproduced as 
RuO$_2$/LaO/MnO$_2$/La$_{0.5}$Ba$_{0.5}$O/RuO$_2$, as a consequence the simulated stoichiometry is La$_{0.75}$Ba$_{0.25}$MnO$_3$ that is very close to the experimental one. 
For Brillouin zone integrations, a 6\AA$^{-1}\times$6\AA$^{-1}\times$6\AA$^{-1}$ k-point grid is used for geometry relaxation and a 12\AA$^{-1}\times$12\AA$^{-1}\times$12\AA$^{-1}$ k-point grid for the determination of the density of states (DOS) and magnetization.  We optimize the internal degrees of freedom by minimizing the total energy to be less than $3\times10^{-5}$\AA eV and the Hellmann-Feynman forces to be less than 20 meV/{\AA}. The Hubbard U effects at the Ti, Ru and Mn sites were included in the LSDA+U approach using the rotational invariant scheme proposed by Liechtenstein\cite{Liechtenstein95}. 
In the literature, the suggested ranges for the Coulomb repulsion are 4-6 eV for the Ti atoms, 0-1 eV for the Ru atoms in the metallic regime\cite{Roy15} and 3-7 eV for the Mn atoms\cite{Keshavarz:2017_PRB,Autieri:2014_NJP}. We have used U=5 eV and J$_H$=0.64 for the Ti atoms, while for the Ru and Mn J$_H$=0.15U is considered.
We have used two sets of values for U$_{Ru}$ and U$_{Mn}$, one with small values of U more appropriate for three-dimensional systems and another with large values of U more appropriate for two-dimensional systems. 
In the first set, we have used the values in the bottom of the range (U$_{Ru}$=0.2 eV, U$_{Mn}$=3 eV) while in the second set we have used values in top of the suggested ranges (U$_{Ru}$=1 eV, U$_{Mn}$=6 eV).
While the first set is closer to the experimental values, the  values of the Coulomb repulsion in the second set helps the convergence of the magnetic moment. Small deviations from these Coulomb repulsion values, do not modify the scenario described in the main text.
The DFT calculation were performed in a collinear approximation, in the realistic case a slightly canting between the spin could be expected.

\section{Octahedral rotations}
DFT results show how the octahedral distortions behaves at the SRO/LBMO interface. 
In the SRO mother compound, the theoretical octahedral distortions are close 160 degrees for both the in plane and out of plane Ru-O-Ru bond angle.
The in-plane M-O-M bond angle in the first layer of STO is constrained to be 180 degrees, therefore the SRO layers closer to the substrate show larger bond angles but they evolve towards the bulk values. 
We can see from the Figure 5(a) in the manuscript that the SRO closer to the LBMO interface with an excess of La is more distorted.
Large octahedral rotations, produces small bandwidth W and small hopping $t$. 
While the magnetic moment increases with the increase of the parameters $U/t$, the critical temperature in itinerant ferromagnets depends on the double exchange that is proportional to $t$.
Therefore, usually the critical temperature and the magnetic moment are inversely correlated in the itinerant ferromagnets.

\section{Density of states and T$_C$ in mean field approximation}
We tried to stabilize the G-type AFM insulating phase, using also U$_{Ru}$=3 eV but the metallic FM phase is always the ground state for the 3|1|3 trilayer. We also tried other antiferromagnetic configurations like A-type and C-type, but the DFT results show that the SRO has the bulk-like 3D itinerant ferromagnetism.
We show the layer resolved Density of states (DOS) for the d-orbitals of the transition metals.
We define the layers with the name of the compound plus an index layer as
STO1, STO2, SRO3, SRO4, SRO5, LBMO6, SRO7, SRO8 and SRO9.
Since the main focus of the paper is the coupling between SRO and LBMO, we plot the local DOS of the SRO4, SRO5, LBMO6, SRO7, SRO8 and SRO9 layers. We observe that all SRO layers are metallic, while the LBMO layer is half-metallic. The Ru and Mn d-states lie in the same energy range producing a strong magnetic and electronic coupling between the Ru and Mn orbitals. The inner layers SRO4 and 
SRO8 present features at -1 eV different from the  other SRO layers. 
In case of interdiffusion of the Mn atoms in the SRO, the system will become AFM insulator. In the case of La interdiffusion in the SRO, the electronic configuration of the Ru will be 4d$^5$ that is expected to be paramagnetic. Therefore, the observed magnetism of the SRO rules out every possible source of noncollinearity and interdiffusion of the LBMO in the SRO.

\begin{figure}[tb]
\includegraphics[width=0.3\textwidth,angle=270]{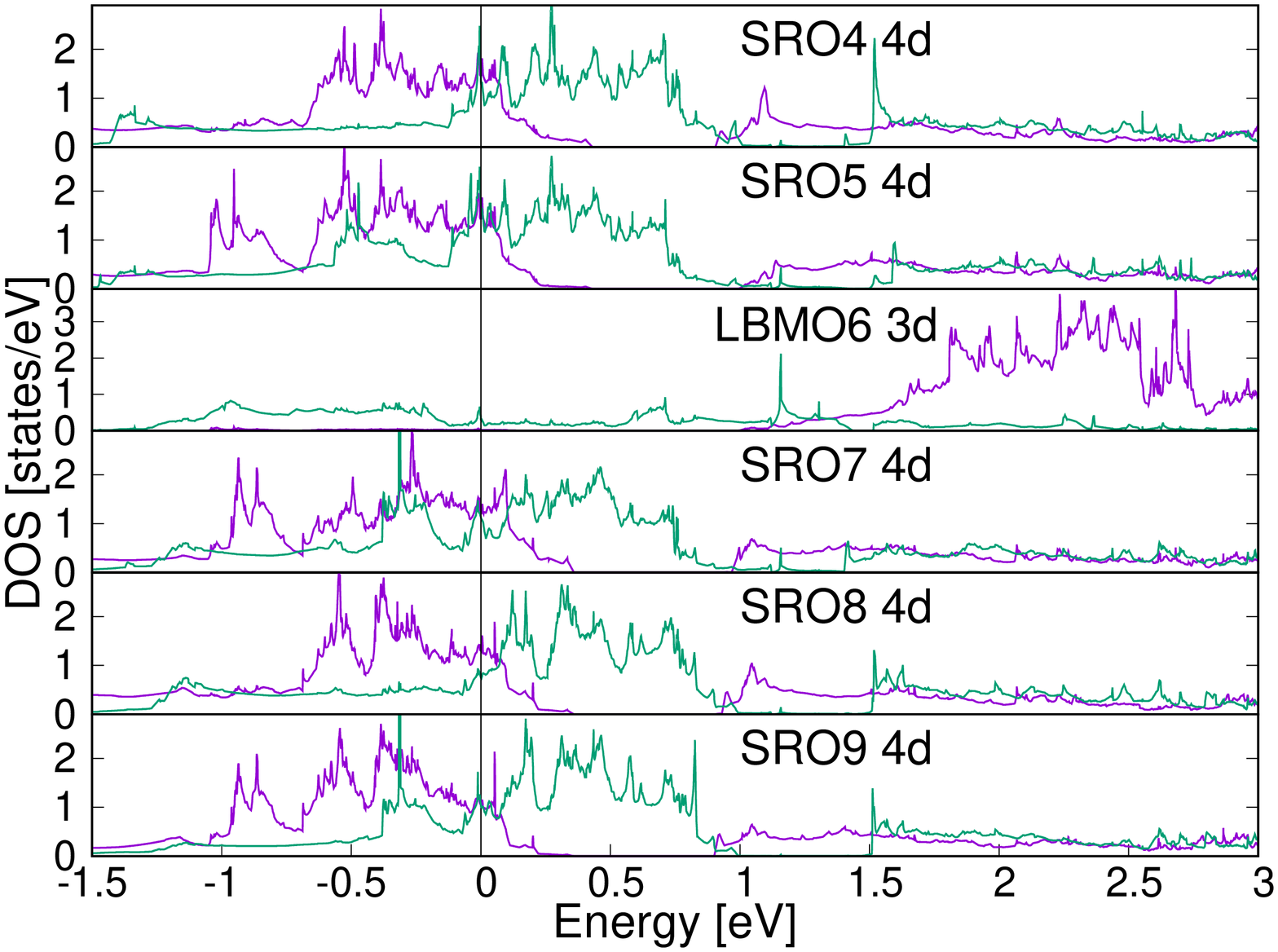}
  \caption{Layer resolved DOS for the d-orbitals of SRO4, SRO5, LBMO6, SRO7, SRO8 and SRO9 in the case of U$_{Ru}$=0.2 eV and U$_{Mn}$=3 eV. Spin up (spin down) states are purple (green). The Fermi level is set to zero.}
  \label{fig:DFT_DOS_U02}
\end{figure}

\begin{figure}[tb]
\includegraphics[width=0.3\textwidth,angle=270]{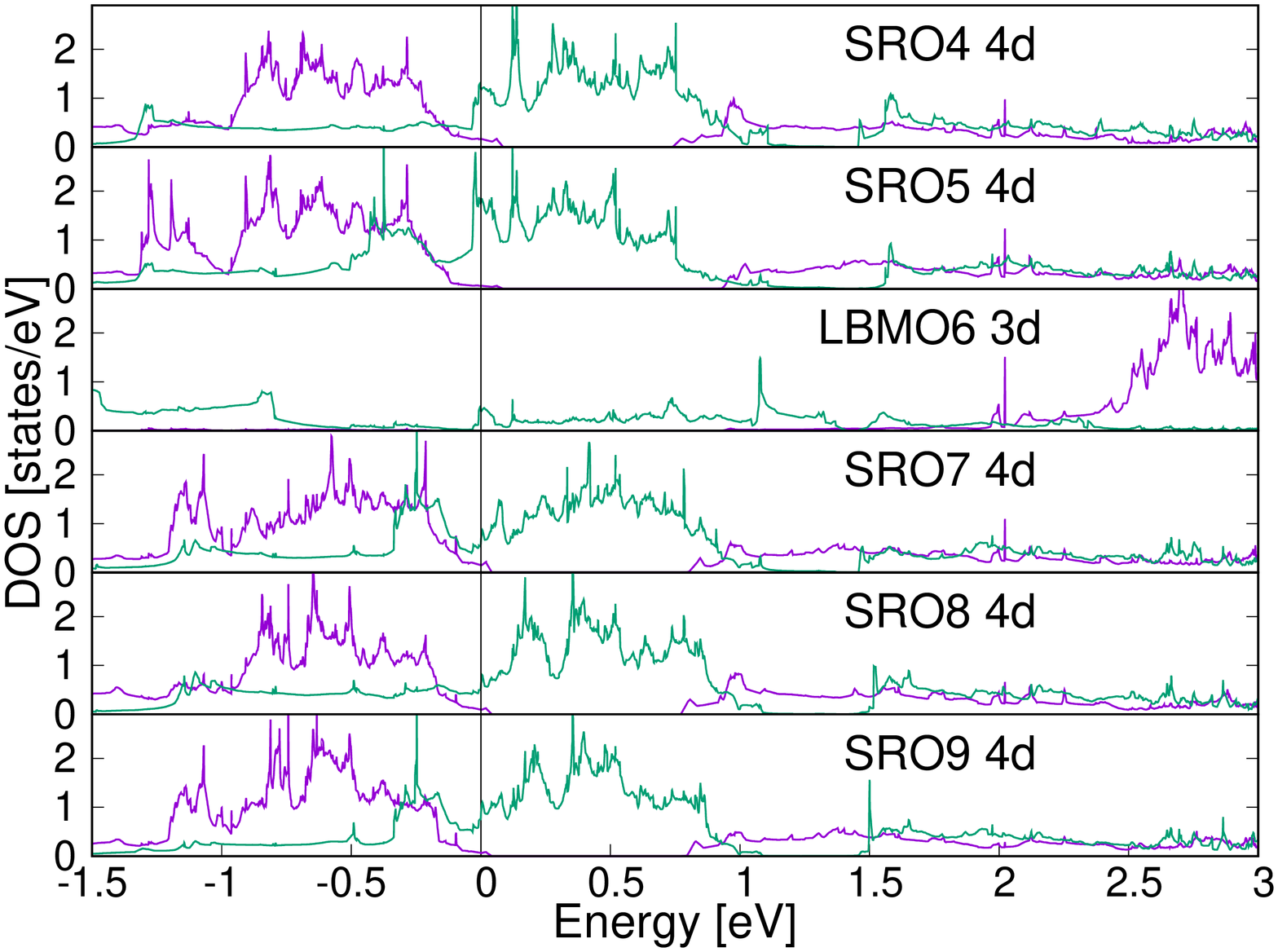}
  \caption{Layer resolved DOS for the d-orbitals of SRO4, SRO5, LBMO6, SRO7, SRO8 and SRO9 in the case of U$_{Ru}$=1 eV and U$_{Mn}$=6 eV. Spin up (spin down) states are purple (green). The Fermi level is set to zero.}
  \label{fig:DFT_DOS_U1}
\end{figure}

The Mn and Ru atoms have different magnetic moment, however, for simplicity we will map both spins in an Heisenberg model with spin S=1.
\begin{equation}
H=\sum_{<i,j>}J_{i,j}\boldsymbol{S_i}{\cdot}\boldsymbol{S_j}
\end{equation}
where $<i,j>$ represent the first neighbor interaction between the atom i and the atom j, where J$_{i,j}$=J$_{j,i}$ but they are both considered in the Hamiltonian.
The mapping on the spin S=1 contribute to overestimate the T$_C$.
Given the crystal structure of the 3|1|3, the Heisenberg Hamiltonian can be described with two main parameters
J$_{Mn,Mn}^{100}$ and J$_{Mn,Ru}^{001}$.
Using the Heisenberg model
we can estimate the T$_C^{MFA}$ in mean field approximation for the Mn atoms using the equation:
\begin{equation}
T_{C,LBMO}^{MFA}=\frac{2}{3k_B}\sum_{<Mn,j>}|J_{Mn,j}|
=\frac{2}{3k_B}(|4J_{Mn,Mn}^{100}|+2|J_{Mn,Ru}^{001}|)
\end{equation}
where j runs over the neighbors of one Mn atom.
Using the first set of U-values we get J$_{Mn,Mn}^{100}$= 23.2 meV and J$_{Mn,Ru}^{001}$= 3.7 meV, while using the second set we get J$_{Mn,Mn}^{100}$= 17.6 meV and J$_{Mn,Ru}^{001}$= 3.2 meV. 
The Mn-Mn magnetic exchange is larger than the Mn-Ru magnetic exchange. The critical temperature in MFA largely overestimates the experimental value as in other TMO perovskites\cite{Ivanov2016} especially in the low connectivity limit and in the case of inequivalent magnetic atoms. However, we are able to capture the qualitative behaviour of the SRO on the LBMO, understand how $J_{Mn,Ru}^{001}$ increases the T$_C$ of the LBMO.
For the first set, the ratio $\frac{{\Delta}T_{C,LBMO}}{T_{C,LBMO}}=\frac{J_{Mn,Ru}^{001}}{2J_{Mn,Mn}^{100}}$ is equal to 0.08  therefore the presence of SRO increases the T$_C$ by 8\% in mean field approximation, while for the second set the increase is 9\%.
We should also stress that the Ru allow to move from a 2D magnetism to 3D magnetism, this has large consequence that are not correctly captured by the mean field approximation.
Also if the SRO is paramagnetic above the critical temperature of the LBMO, the SRO still contains magnetic Ru atoms that couples with the Mn atoms\cite{PhysRevB.91.205116}.

The effect of the SRO on the LBMO is to sustain the critical temperature, the effect of the LBMO on the SRO is to do not destroy the collinearity. This is an additional example on how unreliable is the hypothesis of non-collinearity/skyrmions at the SRO interface.

\section{Details of the semiclassical model} 
The semiclassical model contain a factor 6 because we have 6 SRO layers.
We have used $\frac{K_{Ru}}{M_{Ru}}$=0.60 T (smaller than the bulk value of 2.5 T\cite{Ziese:2010ep}) and $\frac{K_{Mn}}{M_{Ru}}$=0.07 T and J$^{001}_{Ru-Mn}$=+0.70 meV to mimic the experimental results. 
The magnetic coupling used in this model is smaller than the coupling calculated using DFT, the reason of the mismatch could be address to the presence of domains not considered in this simplified semiclassical model.
We used the experimental values 1.3 and 3.4 $\mu_B$ for the magnetization. Once fixed the $\theta_H$ angle, we tuned the magnetic field and calculated the values $\theta_{Ru}$ and $\theta_{Mn}$ for every magnetic field by minimizing the energy.
To convert the unit in the semiclassical model we use the equivalence 1 $\mu_B$=0.05789$\frac{meV}{T}$.
The semiclassical equation show that 
the magnetocrystalline anisotropy in the films is smaller than the bulk, moreover the AFM magnetic coupling between Ru and Mn is in competition with other interactions not included in the model like the DMI. For this reason we need a small J to reproduce the experimental results.

\section{X-ray magnetic circular dichroism measurements}

The XMCD and element specific magnetization measurements were carried out  on the EPFL/PSI X-Treme beamline at the Swiss Light Source, Paul Scherrer Institut, Villigen, Switzerland\cite{Piamonteze:2012jg}. The spectra were acquired by total electron yield detection. The XAS spectra are calculated as the sum of the spectra measured with positive and negative photon helicity. The degree of circular polarization is better than 97\%. The energy resolution was 0.07~eV and 0.1~eV at Ru and Mn edges, respectively.

\section{Sum rule calculations}

\begin{table*}[b]
    \centering
    \begin{tabular}{c|c|c|c|c|c} \hline \hline
    film     &  element & geometry & field (T) & m$_s$ ($\mu_B$) & m$_l$ ($\mu_B$) \\ \hline
    3|1|3 & Mn & NI & 6.8 & 1.7(3) & 0.02(18) \\
    3|1|3 & Mn & NI & 0 & 3.0(6) & 0.07(14) \\
    3|1|3 & Mn & GI & 6.8 & - 0.32(3) & -0.013(3) \\
    3|1|3 & Mn & GI & 0 & 1.9(1) & 0.006(86) \\
    LBMO & Mn & GI & 6.8 & -3.5(6) & 0.037(7) \\
    SRO & Ru & NI & 6.8 & -1.22(4) & -0.15 (2) \\ \hline \hline
    \end{tabular}
    \caption{Sum rule results.}
    \label{tab:SR}
\end{table*}

\begin{figure}[tb]
\includegraphics[width=0.35\textwidth]{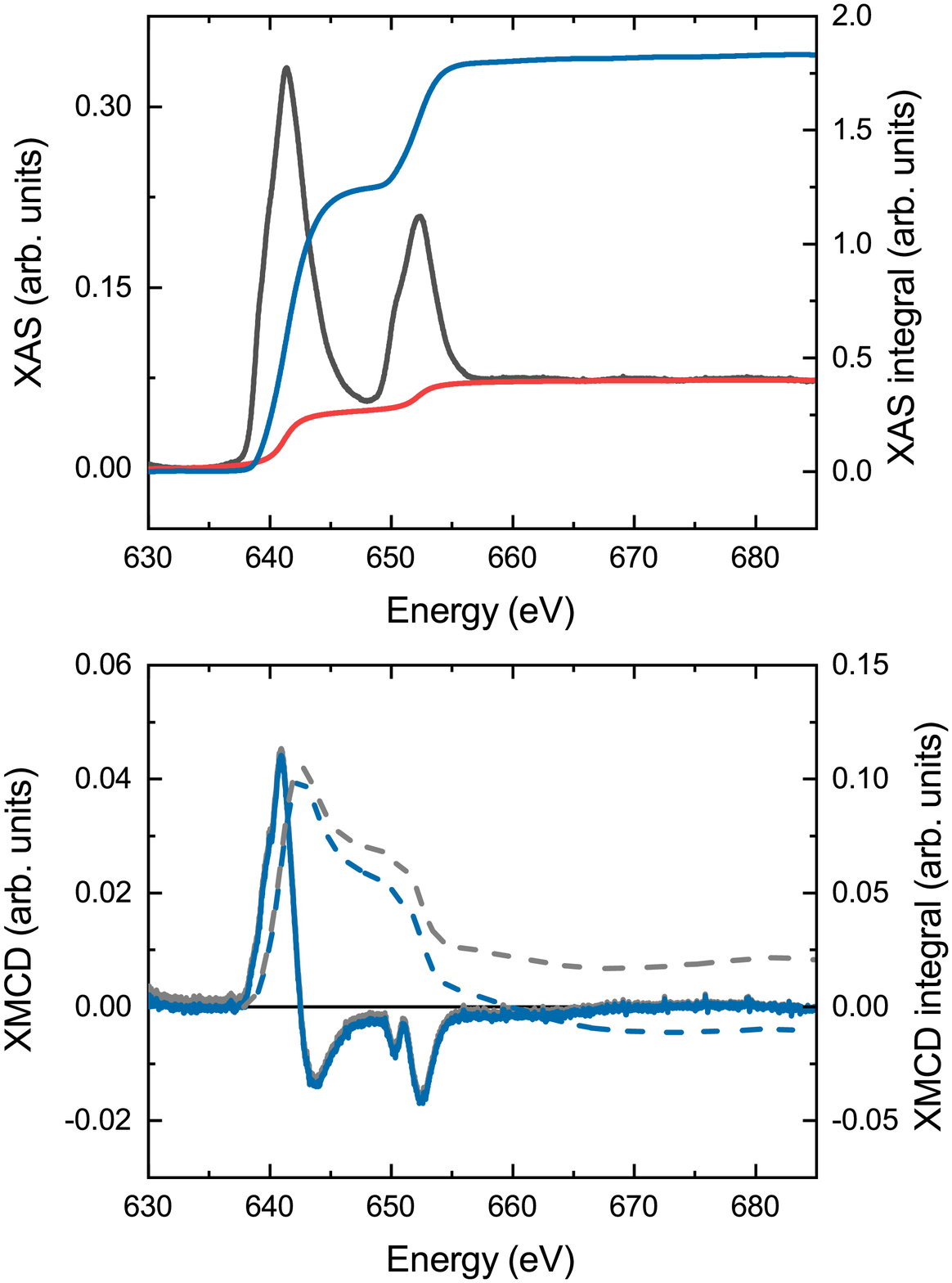}
  \caption{Top: Mn XAS for 3|1|3 trilayer (gray) and step function (red) removed before calculation of XAS integral (blue). Bottom: Mn XMCD for 3|1|3 trilayer (gray/blue, continuous line) at 6.8\,T, 10\,K in normal incidence an respective integral (gray/blue, dashed line).}
  \label{fig:SR_Mn1}
\end{figure}

\begin{figure}[tb]
\includegraphics[width=0.35\textwidth]{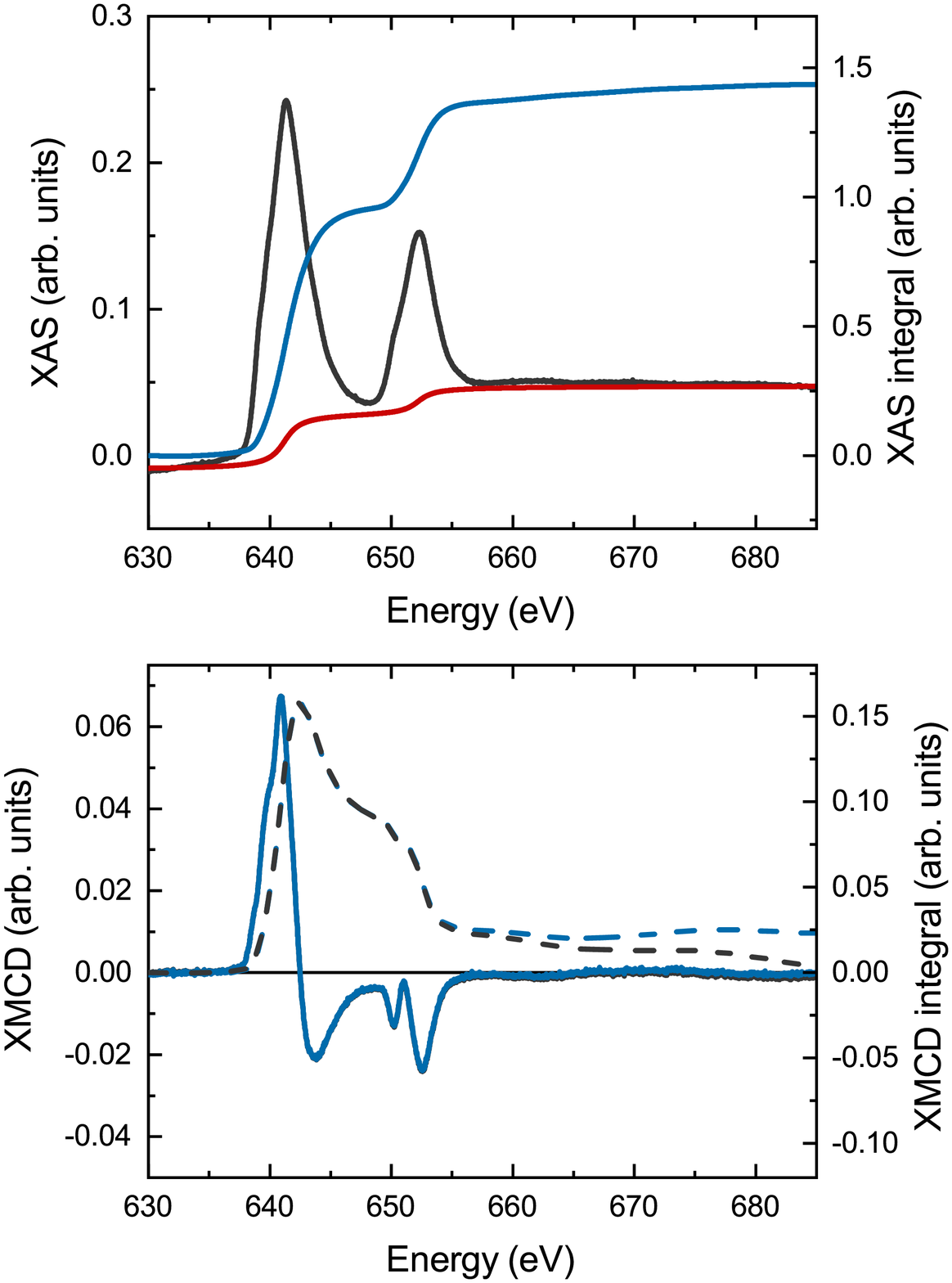}
  \caption{Top: Mn XAS for 3|1|3 trilayer (gray) and step function (red) removed before calculation of XAS integral (blue). Bottom: Mn XMCD for 3|1|3 trilayer (gray/blue, continuous line) at remanence, 10\,K in normal incidence an respective integral (gray/blue, dashed line).}
  \label{fig:SR_Mn2}
\end{figure}

\begin{figure}[tb]
\includegraphics[width=0.35\textwidth]{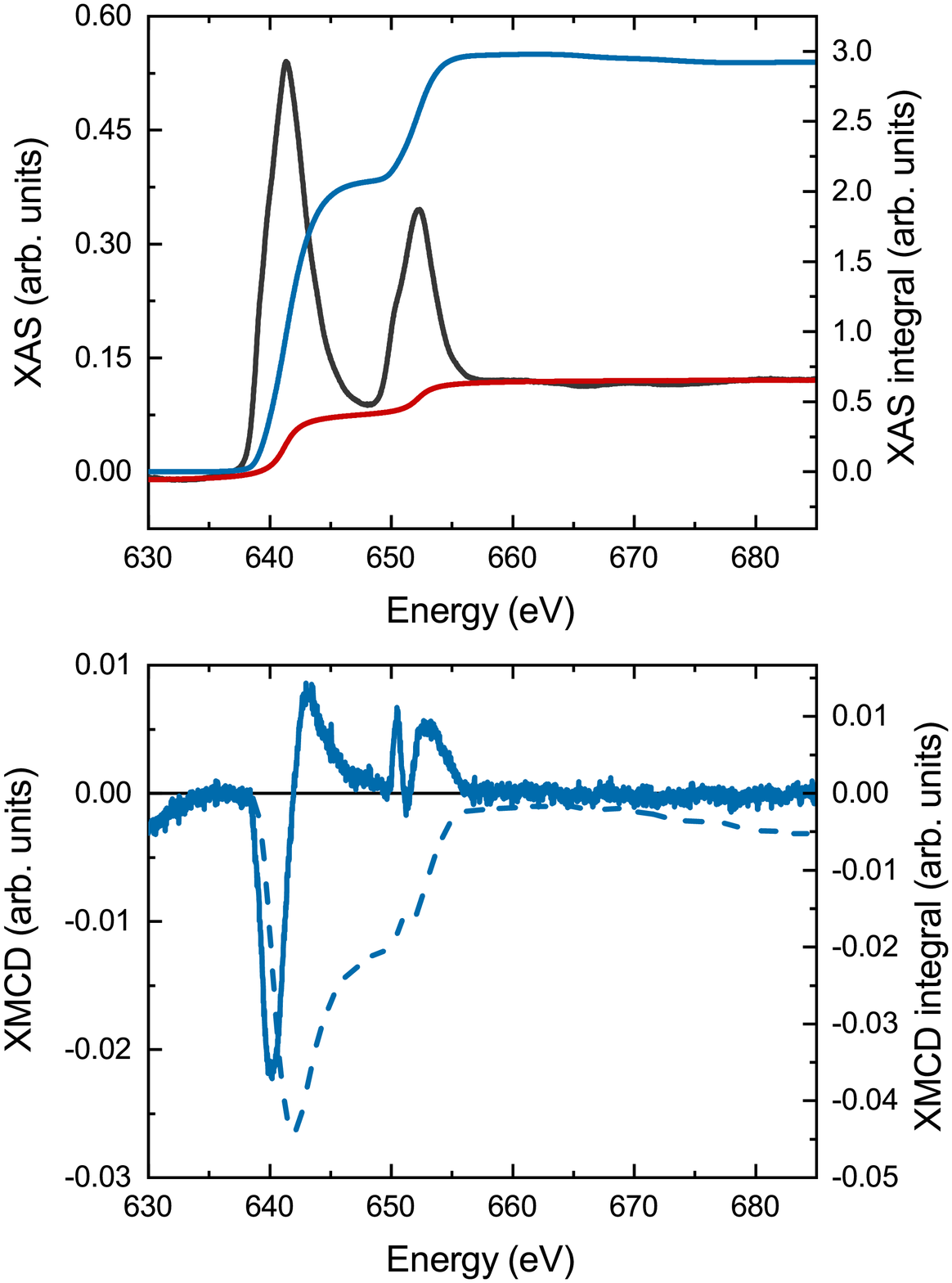}
  \caption{Top: Mn XAS for 3|1|3 trilayer (gray) and step function (red) removed before calculation of XAS integral (blue). Bottom: Mn XMCD for 3|1|3 trilayer (gray/blue, continuous line) at 6.8\,T, 10\,K in grazing incidence an respective integral (gray/blue, dashed line).}
  \label{fig:SR_Mn3}
\end{figure}

\begin{figure}[tb]
\includegraphics[width=0.35\textwidth]{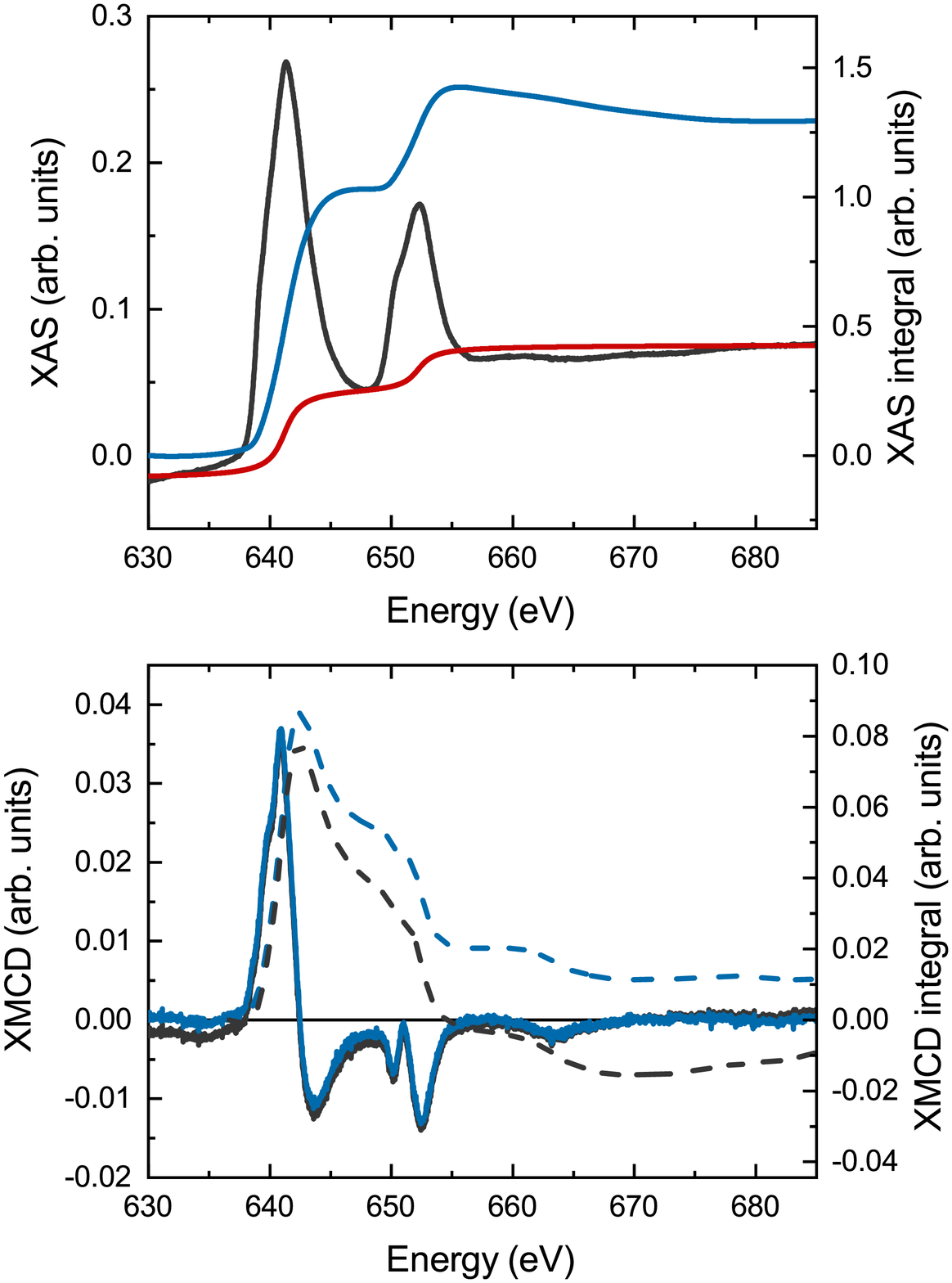}
  \caption{Top: Mn XAS for 3|1|3 trilayer (gray) and step function (red) removed before calculation of XAS integral (blue). Bottom: Mn XMCD for 3|1|3 trilayer (gray/blue, continuous line) at remanence, 10\,K in grazing incidence an respective integral (gray/blue, dashed line).}
  \label{fig:SR_Mn4}
\end{figure}

\begin{figure}[tb]
\includegraphics[width=0.35\textwidth]{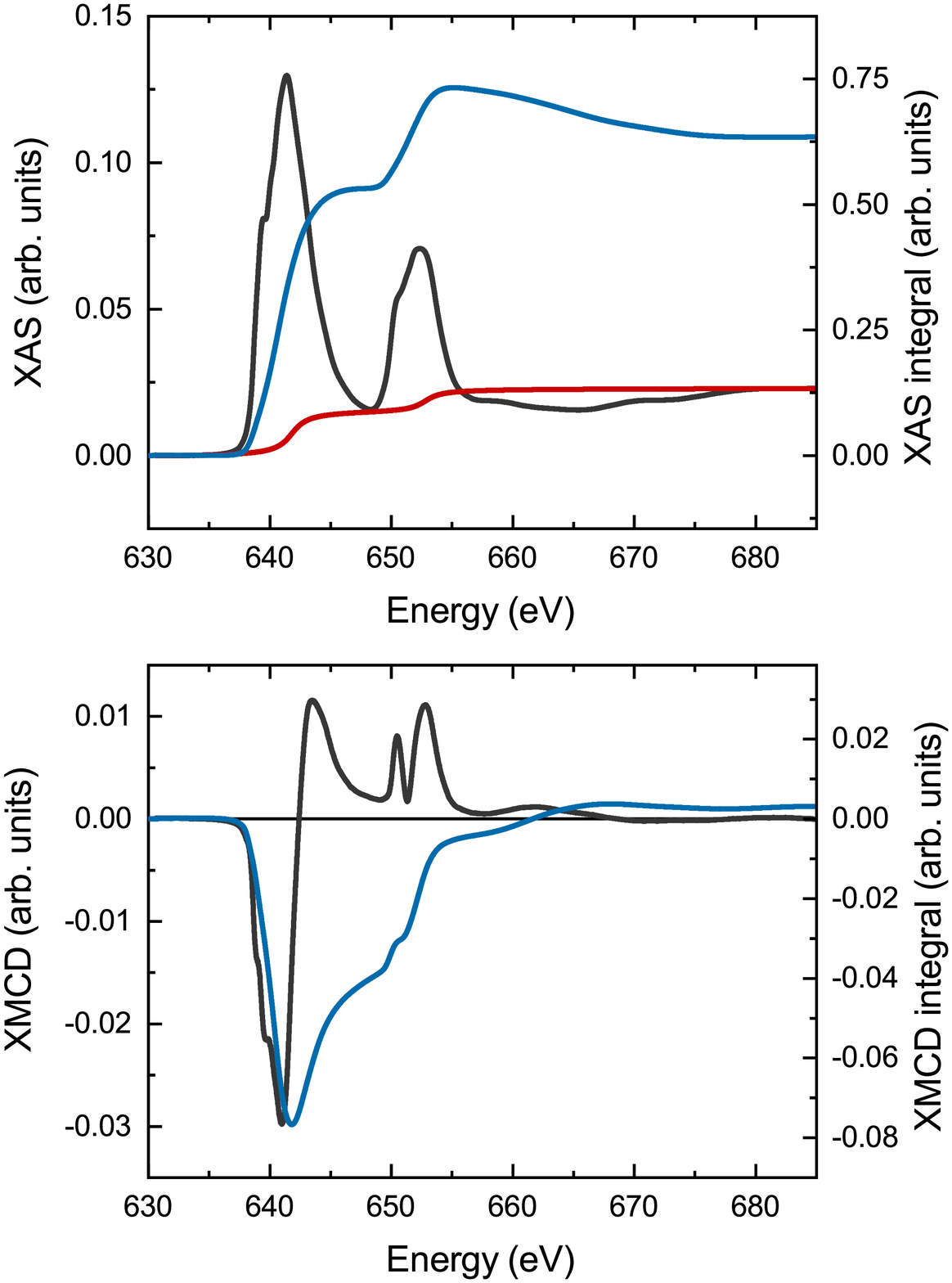}
  \caption{Top: Mn XAS for LBMO single film (gray) and step function (red) removed before calculation of XAS integral (blue). Bottom: Mn XMCD for 3|1|3 trilayer (gray) at remanence, 10\,K in grazing incidence an respective integral (blue).}
  \label{fig:SR_LBMO}
\end{figure}

Figures \ref{fig:SR_Mn1}-\ref{fig:SR_Mn4} show the XAS and XMCD spectra and respective integrals used for the sum rule calculation of  Mn moment of 3|1|3 trilayer in NI and GI in applied field and remanence. The spin sum rules are not exact for Mn due to proximity in energy \cite{Teramura:1996wv} of the L$_3$ and L$_2$ edges. We have calculated the correction factors for the sum rules \cite{Piamonteze:2009id} to be 1.51. The number of holes used was 5.8, obtained from configuration interaction simulation of x-ray photoemission spectroscopy data\cite{Saitoh:1995uc}. As shown in the figures small changes in the baselines of the XMCD can lead to significant changes on the orbital moment. These variations have been taken into account in the calculation of the error bars shown in table \ref{tab:SR}. Figure \ref{fig:SR_LBMO} shows the sum rules analysis for the single film of LBMO. The same correction factor and number of holes was used.

\begin{figure}[tb]
\includegraphics[width=0.35\textwidth]{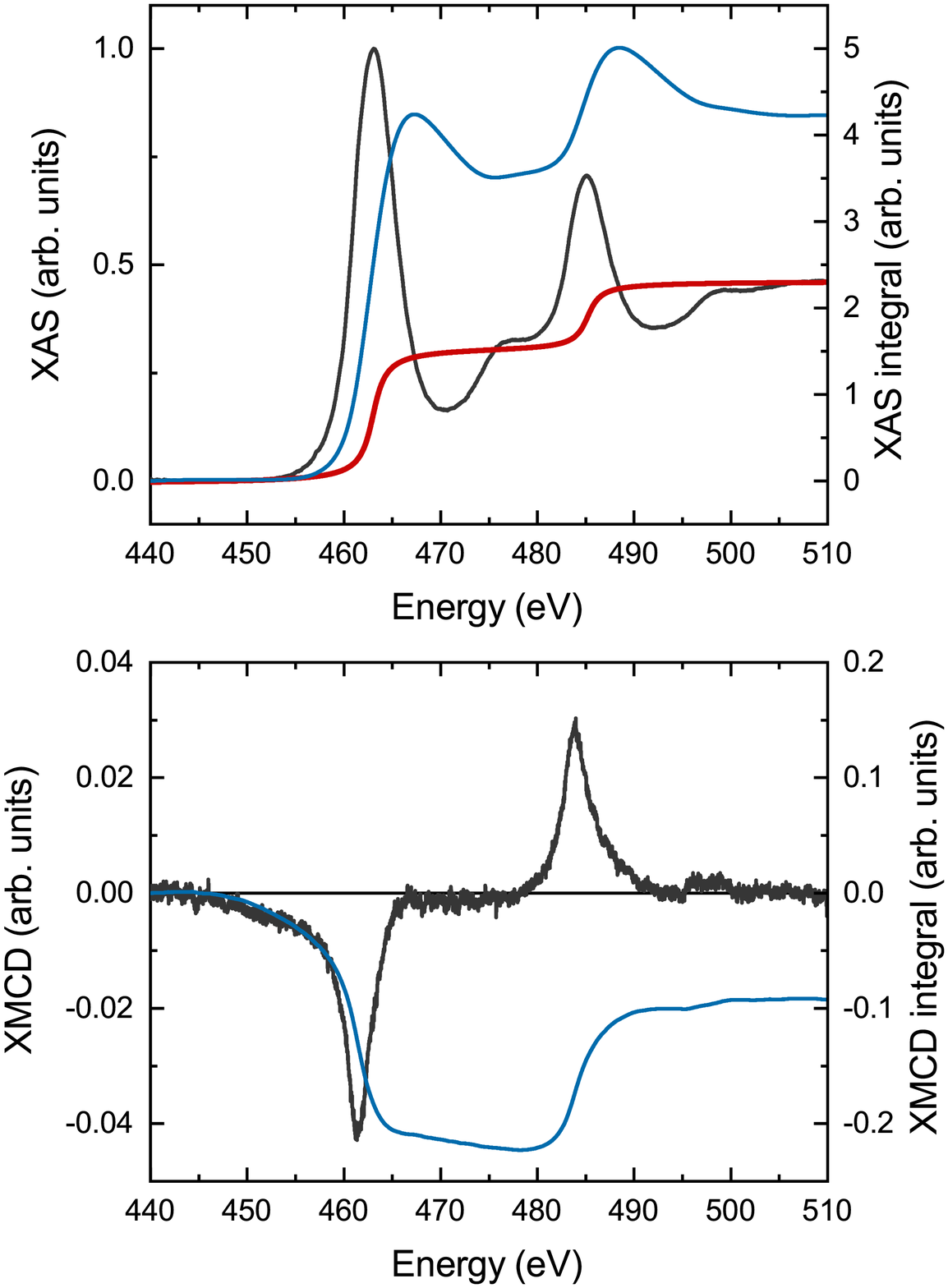}
  \caption{Top: Ru XAS for SRO single film (gray) and step function (red) removed before calculation of XAS integral (blue). Bottom: Ru XMCD for SRO (gray) at remanence, 10\,K in normal incidence an respective integral (blue).}
  \label{fig:SR_SRO}
\end{figure}

There is some discrepancy in the moments obtained by Ru sum rules in the literature \cite{Okamoto:2007dd},\cite{Agrestini:2015km}, \cite{Ishigami:2015ga}. In our case we found particularly challenging to find a good baseline subtraction of the XAS data. This will of course have impact in the calculation of the sum rules later. Here, we have chosen to use the theoretical XAS calculated from the tables by Henke \textit{et. al}\cite{Henke:1993},\cite{Henke:web} to remove the slope coming from Sr and O XAS. After removing this theoretical background, we still removed a line fitted to the pre-edge and obtain the XAS shown in figure \ref{fig:SR_SRO}. We find a non-negligible orbital moment for SRO as also found in \cite{Ishigami:2015ga}. Our m$_l$/m$_s$ is 0.12. The correction factor due to the overlap of the M$_2$ and M$_3$ edges is less significant (we used 1.08) than for Mn since the Ru $3p$ spin-orbit is larger.

The sum rules for the Ru XMCD in 3|1|3 turned out to be difficult due to a large error bar coming from the XMCD baseline in comparison to the total XMCD signal. In order to give an order of magnitude for the moment we have scaled the Ru data from 3|1|3 to the one for SRO. The XMCD are scaled to the XAS shown in figure 1 in the  manuscript. In NI the 3|1|3 Ru XMCD corresponds to 75\% of the one for SRO. The Ru XMCD measured in GI  for 3|1|3 corresponds to 40\% of the one for SRO.

\begin{figure}[tb]
\includegraphics[width=0.4\textwidth]{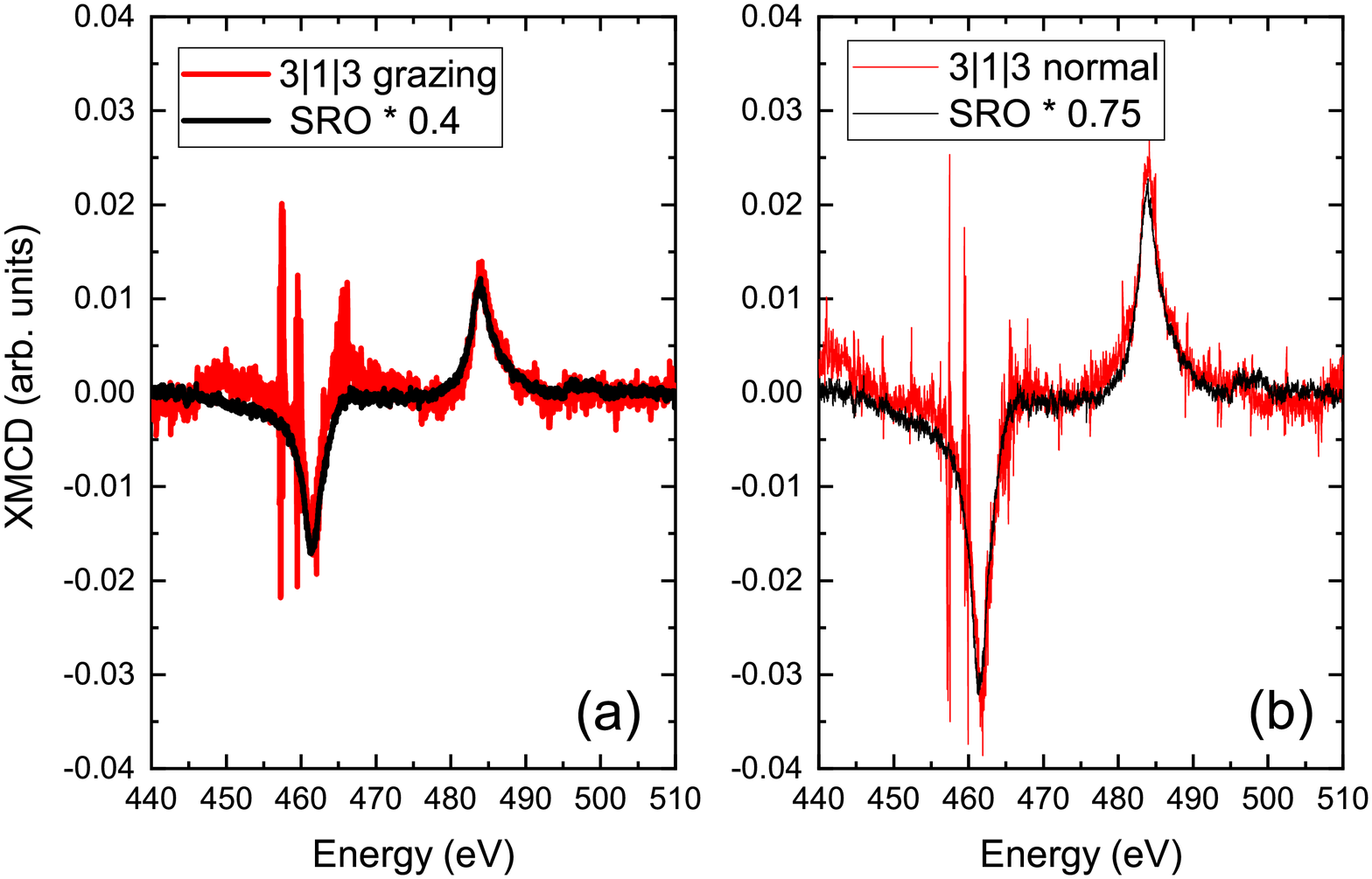}
  \caption{Red curves show XMCD for Ru in 3|1|3 heterostructure in grazing (a) and normal (b) incidence. The  SRO XMCD  scaled to the 3|1|3 data is shown in black.  The data for SRO was measured in normal incidence. All data was measured at 10\,K and 6.8\,T.}
  \label{fig:SR_Ru}
\end{figure}


\clearpage

\bibliography{lbmo_sro}

\end{document}